\documentclass{sig-alternate} 
\usepackage{mathptmx} 

\newcommand{\ignore}[1]{}
\usepackage{fancyhdr}
\usepackage[normalem]{ulem}
\usepackage[hyphens]{url}
\usepackage{microtype}
\usepackage{array}

\usepackage{tikz}
\usetikzlibrary{positioning}
\usepackage[export]{adjustbox}

\usetikzlibrary{patterns,snakes}

\usepackage[bookmarks=true,breaklinks=true,letterpaper=true,colorlinks,linkcolor=black,citecolor=blue,urlcolor=black]{hyperref}

\newcommand{\shrink}{Eliminate vertical white-space}
\newcommand{\vshrink}[1]{
	\ifdefined\shrink 
	\vspace{-#1cm}
	\else
	\vspace{0cm}
	\fi
}

\fancypagestyle{firstpage}{
  \fancyhf{}

  \fancyhead[C]{
  } 
  \fancyfoot[C]{\thepage}
}  

\title{\vspace{-3ex}Temporal-Rate Encoding to Realize Unary Positional Representation in Spiking Neural Systems \vspace{-4ex}} 
\author{Zhenduo Zhai \qquad\qquad\qquad Ismail Akturk \\
	Department of Electrical Engineering and Computer Science\\University of Missouri, Columbia\\
	zz7z9@mail.missouri.edu \qquad akturki@missouri.edu}

\begin{document}
\maketitle
\thispagestyle{firstpage}
\pagestyle{plain}

\begin{abstract}
Unary representation is straightforward, error tolerant and requires simple logic while its latency is a concern. On the other hand, positional representation (like binary) is compact and requires less space, but it is sensitive to errors. A hybrid representation called unary positional encoding reduces the latency of unary computation and length of the encoded stream, thus achieves the compactness of positional representation while preserving the error tolerance of unary encoding. In this paper, we discuss the prospect of unary positional encoding in spiking neural systems by incorporating temporal and rate encoding.

\end{abstract}

\section{Introduction}
%
%
%

There are mainly two classes of encoding approaches in spiking neural systems. The first class is {\em rate encoding} where values are encoded as rates of spikes emitted from a neuron. Assuming a bundle of neurons, each neuron represents a separate value, thus a change in spike rate on a particular neuron would only impact its own value.

The second class of encoding is {\em time encoding} (a.k.a. temporal encoding) where the values are encoded as temporal relationships among the spikes emitted from the neurons in the bundle. Since the values are represented relative to each other, any change in spiking time on a particular neuron may impact the value of all (or some) of the other neurons in the bundle.

Without arguing about which encoding is possibly biologically accurate, we seek to take advantage of both encoding schemes while designing neuromorphic or brain-inspired computing systems. In our wish list, we have an encoding scheme which is compact and has low latency while it is error-tolerant. In traditional systems, we use positional encoding (e.g., binary) where a digit carries a different weight based on its position in the stream of digits. Positional encoding has compact representation but it is sensitive to errors (a single flip or change in a digit may drastically impact the overall value). On the other hand, unary encoding is used in stochastic computing~\cite{stochastic_1969} systems where each digit carries the same amount of information (i.e., position independent), thus more resilient to errors. However, it requires too much space and incurs high overhead. A hybrid scheme called unary positional encoding seems a candidate to fulfill our wish list, as it reduces the latency and length of the encoded stream, thus achieves the compactness of positional representation while preserving the error tolerance of unary encoding. 

In the rest of the paper, we discuss the details of unary positional encoding and its prospect in spiking neural systems by incorporating temporal and rate encoding.

\section{Unary and Positional Encoding}
In biological neural systems, the information possibly be encoded differently compared to how we encode the information in traditional and neuromorphic computing systems. Without loss of generality, we would like to focus on representation of information in {\em traditional} sense (e.g., decimal, or binary values). Assuming information is represented as a stream of digits, each digit has a weight associated w.r.t. its position in the stream. This makes the representation as {\em position-dependent}. In general, a value of n-digit sequence can be calculated as:
\vshrink{0.2}
$$(x_{n-1} x_{n-2}... x_0)_{base} = \sum_{i=0}^{n-1} x_i \times base^i$$
\vshrink{0.4}
where $x_i < base$ (and $0\leq i \leq n-1$). As an example,

\begin{equation*}
\begin{split}
(1101)_2  & = \sum_{i=0}^{3} x_i \times 2^i\\
            &= (1\times 2^0) + (0\times2^1) + (1\times2^2) + (1\times2^3) \\ 
& = (13)_{10}
\end{split}
\end{equation*}

Although, positional representation is compact (in terms of number of digits to be used), it is sensitive to errors and the sensitivity is not uniformly distributed among the positions. For example the high-order digits have higher weights, as a result, they become more sensitive to errors. Assume the 4-bit stream given above, the bit flip on the first bit (least significant) would have an impact in the order of $2^0 = 1$, where as the fourth bit would have an impact in the order of $2^3 = 8$. 
Another drawback of positional representation is that it may require complex logic to design even for simplest operations, such as addition and multiplication (complexity mainly arises due to carry-overs). 

An alternative approach to position-dependent representation is position-independent representation, in which a weight is the same for all the positions (i.e., uniformly distributed). A value can be represented as a steam of digits where the number of digits is equal the value of itself. This representation is called {\em unary encoding} (as it has a base of one -- each digit has a weight of one). The same 4-bit stream given above ($(1101)_2$) can be represented in unary encoding as:
\vshrink{0.1}
\begin{equation*}
	\begin{split}
		(1101)_2 & = (13)_{10} = (1111111111111)_{u} \\
	\end{split}
\end{equation*}
\vshrink{0.5}

Since each digit has the same weight, the unary encoding tends to be more error-tolerant compared to positional encoding. Any erroneous flip on a digit would only yield of error of one which limits the impact of any error occur in the representation. Thus, unary encoding suits very well for the computing paradigms and environments where an error (or noise) is not an exception but a norm (e.g., neuromorphic computing). 

Another benefit of unary encoding is the simplicity of logic to be designed for basic operations. Consider an addition in unary representation: it boils down to concatenation of two unary values to be added. Similarly, multiplication boils down to concatenating as many copies of the multiplicand as the value of the multiplier. Although its simplicity, unary encoding and operation on unary values suffer from high latency and storage. It is evident that as the value to be represented increases, more unary digits are necessary, and an operation on unary values may incur high latency due to the number of digits to deal with (compared to positional representation). If we would like to generalize the storage complexity of unary and positional representation in terms of length of the stream (i.e., number of digits), we can derive that
with n-digit stream in $base_x$, we can represent ${base_x}^n$ unique values. On the other hand, to represent the same number of unique values (${base_x}^n$) in unary, we need ${base_x}^n$ digits. 
Table~\ref{table:length-comparison} shows the number of digits needed to represent n-digit positional value in unary. It is clear that the number of digits (thus the storage overhead) in unary encoding increases exponentially.

\begin{table}[!t]
	\caption{Number of digits needed to represent n-digit position-dependent value in unary encoding.}
	\setlength\extrarowheight{2pt}
	\centering
	\begin{tabular}{l|l|l}
	\hline
						& Unary		& Example   \\
	\hline
	Binary	&  $2^n$	&  $(1101)_2$ ->  $\underbrace{(1111111111111)_u}_\text{13 digits $\approx  2^4$}$  \\
	\hline
	Decimal 	&  $10^n$	&  $(9876)_{10}$ ->  $\underbrace{(1111...1111)_u}_\text{9876 digits $\approx  10^4$}$ \\
	\hline
	Base x				&  $x^n$	&  $(a_{n-1} ... a_0)_{x}$ ->  $\underbrace{(1111...1111)_u}_\text{$x^n$ digits}$ \\
	\hline
	\hline
	\end{tabular}
\vshrink{0.4}
\label{table:length-comparison}
\end{table}

So far, we have briefly discussed about the pros/cons of positional and unary encoding. 
Now, the question is "can we have the best of both worlds?", namely the simplicity and error-tolerance of unary encoding, and compactness (and performance) of positional encoding.

Hagen and Riedel~\cite{unary-positional} introduced a unary positional representation that is a hybrid approach to achieve the best of both worlds. In unary positional representation, each position has a separate stream of digits (instead of a single digit in traditional positional encoding). The value of each stream is represented as unary, so the value of a stream is the number of ones it contains. Since each stream has a unique weight w.r.t. its location in the set of streams, the overall value can be calculated as weighted sum of each stream. That is:
\vshrink{0.1}
\begin{equation*}
	\vshrink{0.2}
	\begin{split}
& \sum_{i=0}^{k-1} ( \sum_{j=0}^{n-1} x_{i,j} ) \times base^i \\
&  = (\underbrace{ \underbrace{x_{k-1, n-1} x_{k-1, n-2}... x_{k-1, 0}}_\text{n digits} \quad ... \quad \underbrace{x_{0, n-1} x_{0, n-2}... x_{0, 0}}_\text{n digits} }_\text{k positions})_{base} 
	\end{split}
\end{equation*}

\noindent where n is the base of unary positional representation, thus the length of each stream, and k is the number of streams (in positional representation -- weighted streams). The total number of unique values that can be represented in unary positional encoding is $n^k$ (where n is assumed to be a power of 2 for easy conversion from binary to unary positional representation). As an example of unary positional representation in base 8 (i.e., n=8), assume we have a decimal value of 355 that can be represented as $(101100011)_2$ in binary. Then, the number of streams (i.e., k) becomes:
\begin{equation*}
	\begin{split}
		k & = \text{(\# of digits in base 2)} / log(n)_2  \quad  (n = 8)\\
		& = 9 / 3 = 3 \quad \text{(number of streams)}
	\end{split}
\end{equation*}

Once both k and n are known, we can convert $(101100011)_2$ into unary positional representation:
\begin{equation*}
	\vshrink{0.2}
	\begin{split}
		(101100011)_2 & = \fontsize{8.5}{0}(\overbrace{\underbrace{\underbrace{1}_{2^2} \underbrace{0}_{2^1}\underbrace{1}_{2^0}}_\text{convert to unary} \quad \underbrace{\underbrace{1}_{2^2} \underbrace{0}_{2^1}\underbrace{0}_{2^0}}_\text{convert to unary} \quad \underbrace{\underbrace{0}_{2^2} \underbrace{1}_{2^1}\underbrace{1}_{2^0}}_\text{convert to unary}}^\text{k = 3 streams} )_2 \\
		& = \;\;\underbrace{\text{0 1111 00 1}}_\text{n = 8} \quad\: \underbrace{\text{0 1111 00 0}}_\text{n = 8} \quad\: \underbrace{\text{0 0000 11 1}}_\text{n = 8}\\
		&= (0 1111 00 1 \quad 0 1111 00 0 \quad 0 0000 11 1)_{u8}
	\end{split}
\vshrink{0.1}
\end{equation*}
Conversion from binary to the unary positional representation is performed via filling the bits in each stream in expanding groups, considering the original binary bits. The least significant bit in binary is represented by only one bit ($2^0$) in unary positional encoding, then the next least significant bit in binary is represented by two bits ($2^1$), then the third least significant bit in the binary is represented by four bits ($2^2$), and so on. The most significant bit in binary is represented by $2^{m-1}$ bits in unary positional encoding, where m is the number of bits in each binary stream (i.e., total number of bits in binary divided by the number of streams -- k). If the binary bit is one, then the group is filled with ones; otherwise it is filled with zeroes as shown above. The last digit in each stream is always set to zero. Conversion from unary positional to decimal can be performed by splitting the digits into streams of 8 (since base is 8), then assign weight to each stream w.r.t. its location. Finally, by multiplying the unary value in each stream with its weight and adding them together would give the result in decimal.
\begin{equation*}
	\vshrink{0.1}
	\begin{split}
		&\qquad (\underbrace{\fontsize{11}{0}\text{$0 1111 00 1$}}_\text{n = 8} \quad\: \underbrace{\fontsize{11}{0}\text{$0 1111 00 0$}}_\text{n = 8} \quad\: \underbrace{\fontsize{11}{0}\text{$0 0000 11 1$}}_\text{n = 8} )_{u8} \\
		& = \qquad 5\times8^2 \quad + \quad 4\times8^1 \quad  + \quad 3\times8^0\\
		& = \qquad 320 \qquad + \quad 32 \qquad\;\;  + \quad 3 \\
		& = \qquad 355
	\end{split}
\vshrink{0.1}
\end{equation*}

As seen example above, unary positional encoding is more compact as opposed to unary encoding, and less sensitive to errors compared to positional encoding. A single error on a unary positional encoding may impact the value by a smaller amount (up to $n^{k-1}$) compared to positional representation (where the impact can be up to $2^{(log(n)_2\times k)-1}$ -- i.e., the most significant bit flips). Overall, unary positional encoding provides a good compromise between error-tolerance and compactness.

Next, we would like to see how unary positional encoding can be incorporated into temporal and rate encoding in spiking neural systems where we would like to have compactness, low latency and error-tolerance.

\section{Temporal-Rate Encoding}
\begin{figure}[h]
	\centering
\begin{tikzpicture}
\node (fix-begin) at (-1,0) {};
\node (fix-end) at (2.5,0) {};
\node[inner sep=0pt] (unary-rate) at (0,0)
{\includegraphics[width=0.7\columnwidth]{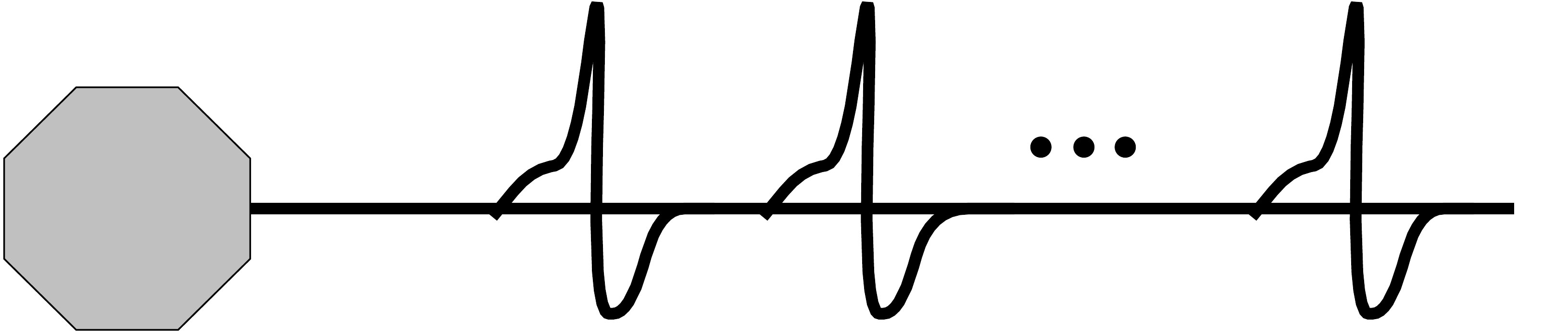}};
\draw [
thick,
decoration={
	brace,
	mirror,
	raise=0.5cm
},
decorate
] (fix-begin.south) -- (fix-end.south)
node [pos=0.5,anchor=north,yshift=-0.55cm] {r = 355 spikes};
\end{tikzpicture}
\vshrink{0.4}
	\caption{A value of $(101100011)_2$ = $(355)_{10}$ encoded as a unary in rate encoding. A neuron spikes as many times as the value to be represented.}	
	\label{fig:unary-rate}
\end{figure}
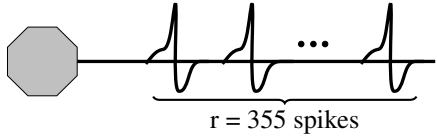

A naive approach to represent a value as unary in spiking neural system would be to use rate encoding. A value can be encoded with as many spikes as there are unary digits. Figure~\ref{fig:unary-rate} illustrates how a value of $(101100011)_2$ = $(355)_{10}$ can be encoded as unary by using rate encoding. Since $(355)_{10}$ requires 355 digits in unary encoding, there are 355 spikes needed. Although a single neuron would suffice to encode the value, it takes too much time to finish the encoding. 
Despite its error-tolerance (missing or having an extra spike impacts the value by -/+ 1/r, where  r is the number of unary digits to be encoded), in many applications an excessive latency
may easily become prohibitive.

\vshrink{0.2}

\begin{figure}[h]
	\centering
	\begin{tikzpicture}
	\node (fix-begin) at (-1,-5.3) {};
	\node (fix-end) at (4.2,-5.3) {};
	
	\node (arrow-begin) at (-1,-5.2) {};
	\node (arrow-end) at (2,-5.2) {};
	
	\node[inner sep=0pt] (unary-rate) at (1,0)
	{\includegraphics[width=0.7\columnwidth]{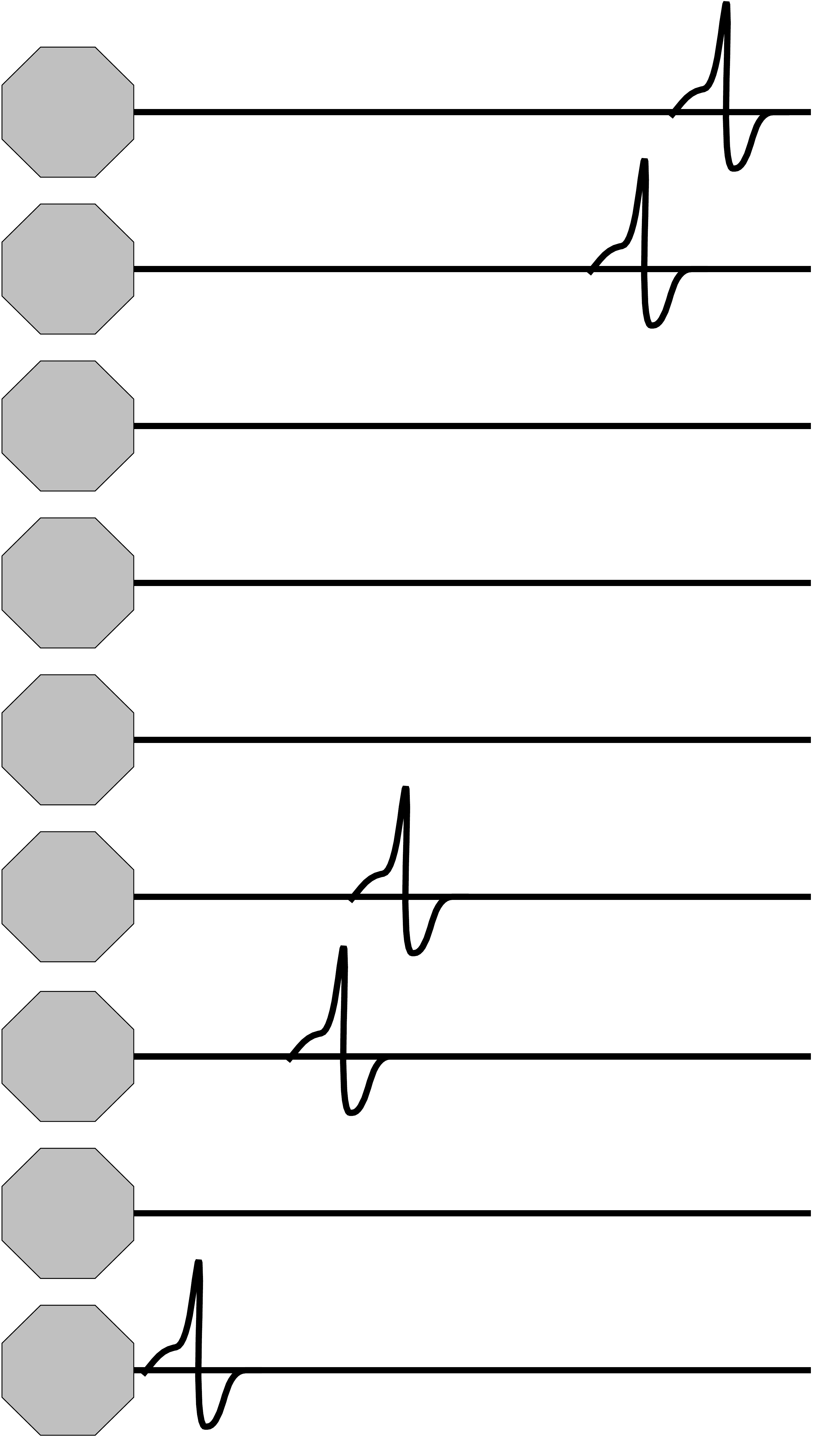}};
	
	\node [] (n1) at (4.5,4.3)   {$t8 = 2^0$};
	\node [] (n1) at (4.5,3.2) 	 {$t7 =2^1$};
	\node [] (n1) at (4.5,2.1) 	 {$t6 =2^2$};
	\node [] (n1) at (4.5,1.0) 	 {$t5 =2^3$};
	\node [] (n1) at (4.5,-0.1) 	 {$t4 =2^4$};
	\node [] (n1) at (4.5,-1.2) 	 {$t3 =2^5$};
	\node [] (n1) at (4.5,-2.35) 	 {$t2 =2^6$};
	\node [] (n1) at (4.5,-3.45) 	 {$t1 =2^7$};
	\node [] (n1) at (4.5,-4.55) 	 {$t0 =2^8$};
	
	\draw [thick, ->] (arrow-begin) edge (arrow-end)
	node [pos=0.5,anchor=north,yshift=-4.75cm,xshift=0.6cm] {time};
	
	\draw [
	thick,
	decoration={
		brace,
		mirror,
	},
	decorate
	] (fix-begin.south) -- (fix-end.south)
	node [pos=0.5,anchor=north,yshift=-0.2cm] {
		$\begin{aligned}
		&(2^0\times 1) + (2^1\times 1) + (2^2\times 0) + (2^3\times 0) + (2^4\times 0) \\
		+ &(2^5\times 1) + (2^6\times 1) + (2^7\times 0) + (2^8\times 1)\\
		= &(355)_{10}
		\end{aligned}$};
	
	\end{tikzpicture}
\vshrink{0.4}
	\caption{A value of $(101100011)_2$ = $(355)_{10}$ encoded as positional (binary) in temporal encoding. Each neuron is assigned a weight relative to its spiking time. 
	}	
	\label{fig:positional-temporal}
	\vshrink{0.5}
\end{figure}

As discussed earlier, positional encoding is compact so it can reduce the latency (compared to unary). Positional encoding can be incorporated in spiking neural systems by exploiting temporal encoding. 
In temporal encoding, the relative spiking times of neurons in a bundle are used as a way to encode a value. To expose positional representation in temporal encoding, weights can be assigned to each neuron w.r.t. their relative spiking time. 
A neuron that spikes first gets the highest weight. Figure~\ref{fig:positional-temporal} illustrates a positional representation of value of $(101100011)_2$ = $(355)_{10}$ in temporal encoding. Since the base for the representation is chosen as 2 (i.e., binary), the weights are assigned to neurons as a power of 2 (the earliest neuron gets a weight of $2^8$ and the last neuron gets a weight of $2^0$). The value can be calculated as a sum of weights assigned to neurons. If a neuron does not spike, it does not contribute to the sum.

Despite its compactness, compared to rate encoding for unary representation, temporal encoding has two main drawbacks for positional values. The first one is error-tolerance. Since the neurons do not carry the same weight, a relative shift in spiking time would render considerable (or limited) change in the value, depending on which neuron experiences the error. The determination of the impact of an error 
on the value is not straightforward in temporal encoding, compared to traditional positional encoding. In traditional positional encoding the impact is in the order of the corresponding weight of the erroneous digit. In temporal encoding, the weights of all (or some) of the neurons may change depending on how much the spiking time dislocated relative to others.

The second drawback of temporal encoding for positional values is low bandwidth utilization. Although each neuron may spike multiple times, only the very first spike matters to assign the weights (rest of the spikes has no impact). Thus, the effective utilization of total spiking capacity of a bundle is limited. Also, the size of a bundle grows as the number of positions increases, since more weights need to be assigned to separate neurons. 
Notice that the bundle size is fixed in rate encoding for unary representation (i.e., a single neuron would suffice to encode any arbitrarily long stream, although it may take longer).
As bundle size increases the latency may also increase since each neuron should spike at different time to have unique weights.  

\begin{figure}[h]
	\centering
	\begin{tikzpicture}
	\node (fix-begin) at (-0.5,-2.3) {};
	\node (fix-end) at (4.0,-2.3) {};
	
	\node (arrow-begin) at (-0.5,-2.2) {};
	\node (arrow-end) at (2,-2.2) {};
	
	\node[inner sep=0pt] (unary-rate) at (1,0)
	{\includegraphics[width=0.7\columnwidth]{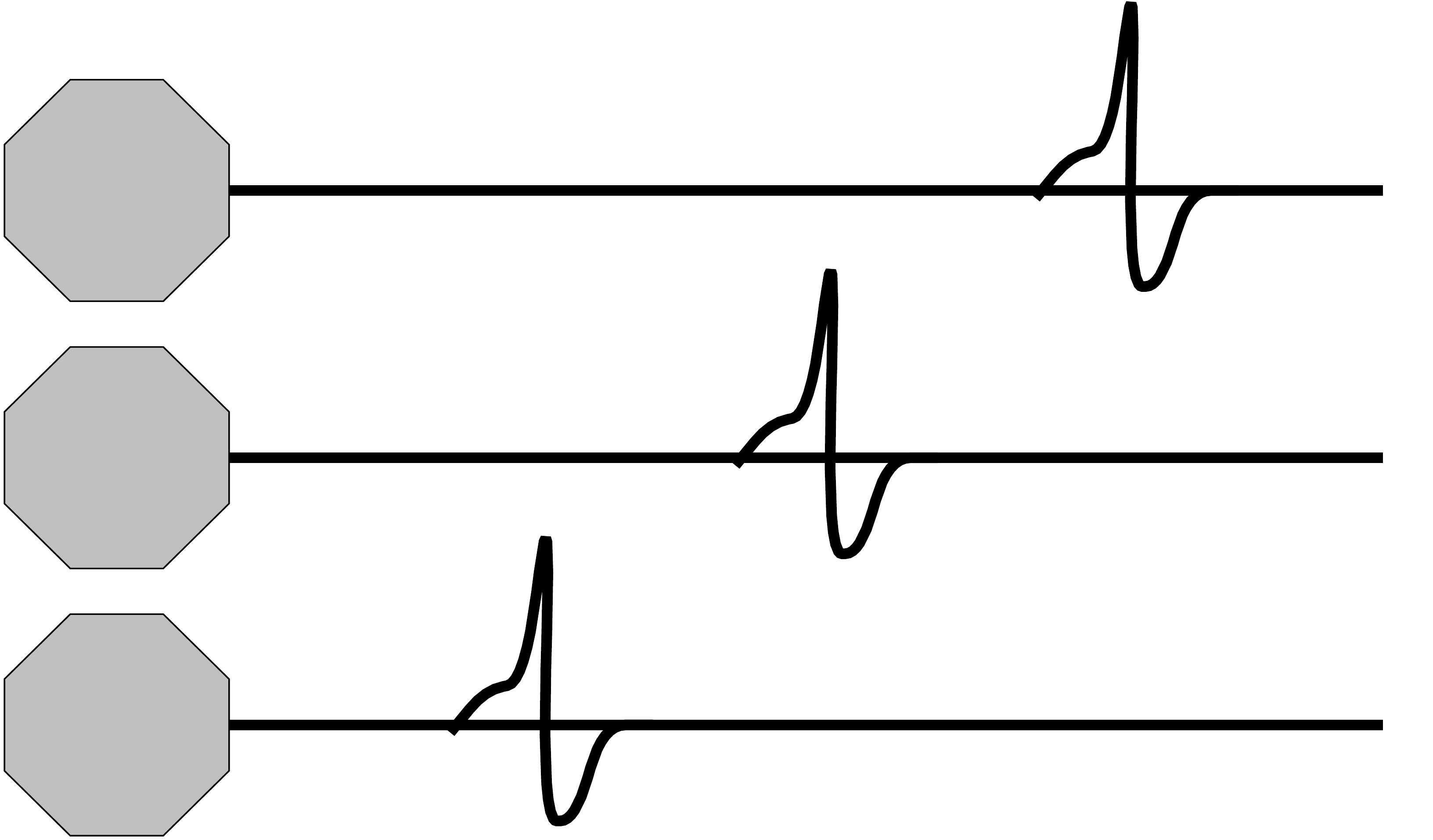}};
	
	\node [] (n1) at (4.5,1)   		 {$t2 = 8^0$};
	\node [] (n1) at (4.5,-0.1) 	 {$t1 =8^1$};
	\node [] (n1) at (4.5,-1.2) 	 {$t0 =8^2$};

	\draw [thick, ->] (arrow-begin) edge (arrow-end)
	node [pos=0.5,anchor=north,yshift=-1.75cm,xshift=0.6cm] {time};
	
	\draw [
	thick,
	decoration={
		brace,
		mirror,
	},
	decorate
	] (fix-begin.south) -- (fix-end.south)
	node [pos=0.5,anchor=north,yshift=-0.2cm] {
		$\begin{aligned}
		&(111)_{t8}\\
		= &(8^2\times 1) + (8^1\times 1) + (8^0\times 1)\\
		= &64 + 8 + 1 = (73)_{10}\qquad \text{(which is > $8^2$  \& < $8^3$)}
		\end{aligned}$};
	
	\end{tikzpicture}
\vshrink{0.4}
	\caption{A value of $(101100011)_2$ = $(355)_{10}$ encoded as positional (base 8) in temporal encoding. Each neuron is assigned a weight relative to its spiking time. The temporal encoding in higher bases have lower resolution. 
	}	
	\vshrink{0.7}
	\label{fig:positional-temporal-base8}
\end{figure}
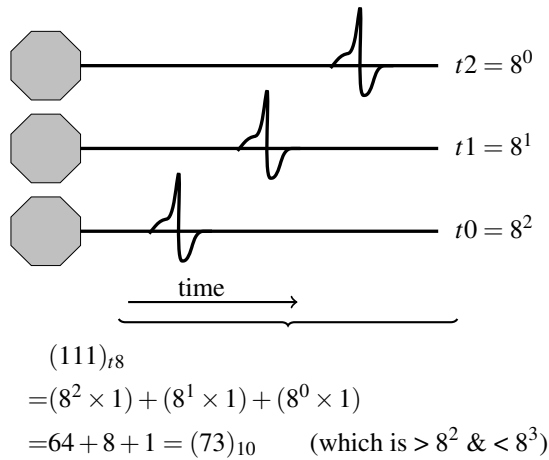

A bigger base can be used to represent a value in temporal encoding to reduce the number of neurons needed in the bundle. This can help to reduce the latency and un-utilized bandwidth.
However, this would result in lower resolution compared to binary. Certain values cannot be represented in full accuracy (as opposed to binary), and distinct values may have the same representation in higher base when encoded as temporal. Figure~\ref{fig:positional-temporal-base8} illustrates a temporal encoding of a value of $(101100011)_2$ in base 8. Each neuron is assigned a weight of power of 8 based on relative spiking time. 

Using temporal encoding for higher bases (e.g., base 8 in this example) has additional challenges besides lower resolution. One is that there is no way to tell if one value is bigger or smaller than the other one by just looking at their representations in base 8 temporal encoding, if their highest weights are the same. For example, assume a stream of digits of $(100)_{t8}$ and $(111)_{t8}$ that represent two separate values in temporal encoding in base 8 (subscript ${t8}$ indicates temporal encoding in base 8). The decimal values of them are $ (100)_{t8} = (64)_{10}$, and $ (111)_{t8} = (73)_{10}$.


In ordinary (not temporal) base 8 a stream of $(111)_8 > (100)_8$. So, we expect that for any arbitrary values $a$ and $b$ in decimal where $a > b$, the temporal representation of them in base 8 ($(A)_{t8}$ and $(B)_{t8}$, respectively) would be $(A)_{t8} > (B)_{t8}$, as well. However, it may not be the case.
Following the example given above, let's look at other possible decimal values that could be represented as $(100)_{t8}$ and $(111)_{t8}$ in base 8 temporal encoding. Among many numbers, we picked the following ones for making our case clear.
In base 8 temporal encoding, $(137)_{10}$, $(145)_{10}$ and $(217)_{10}$ would all be encoded as $(111)_{t8}$, which are smaller than $(256)_{10}$, $(384)_{10}$ and $(448)_{10}$ that would be encoded as $(100)_{t8}$. So, temporal encoding in higher bases may not be reliable to make a comparison unless the number of digits (thus the highest weight) are different (the values whose highest weights are larger than the others' are guaranteed to be bigger). This is mainly because the multipliers of weights cannot be expressed in temporal encoding for higher bases (i.e., $x_i$  can be 0 or 1 in temporal encoding, as opposed to ordinary base representation where $0 \leq x_i \leq base-1$ for a value of $\sum_{i=0}^{n-1} x_{i} \times base^i$).


The famous experiment conducted by Thorpe and Imbert~\cite{Thorpe89} suggests that the very first spike carries the most of the information. This seems inline with our observation for temporal coding, even for higher bases. Assuming a classification or clustering operation to be performed for a set of stimuli, they can be classified or clustered with high accuracy based on their highest weight (i.e., very first spike). For further classification or clustering after the initial clustering, there has to be more information which could not be carried just by the spike time. This is where spiking rate comes into the picture. Although initial spike timing carries the most of the information, more may be needed for better accuracy or higher resolution which can be carried out by the rate of spikes (think of encoded values with rates corresponds to coefficients of weights in unary positional representation). As spiking rate carries less information (compared to timing), the error (+ or -) on rate would have minimal impact on the overall accuracy which makes the encoding more error-tolerant. Thus, temporal and rate encoding seem to be complementary to achieve compactness (so lower latency) and error tolerance in spiking neural systems, similar to unary positional representation in traditional systems.

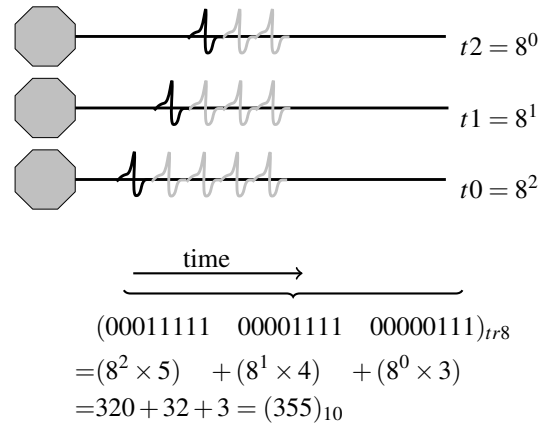
\begin{figure}[h]
	\vshrink{0.3}
	\centering
	\begin{tikzpicture}
	\node (fix-begin) at (-0.5,-2.3) {};
	\node (fix-end) at (4.0,-2.3) {};
	
	\node (arrow-begin) at (-0.5,-2.2) {};
	\node (arrow-end) at (2,-2.2) {};
	
	\node[inner sep=0pt] (unary-rate) at (1,0)
	{\includegraphics[width=0.7\columnwidth]{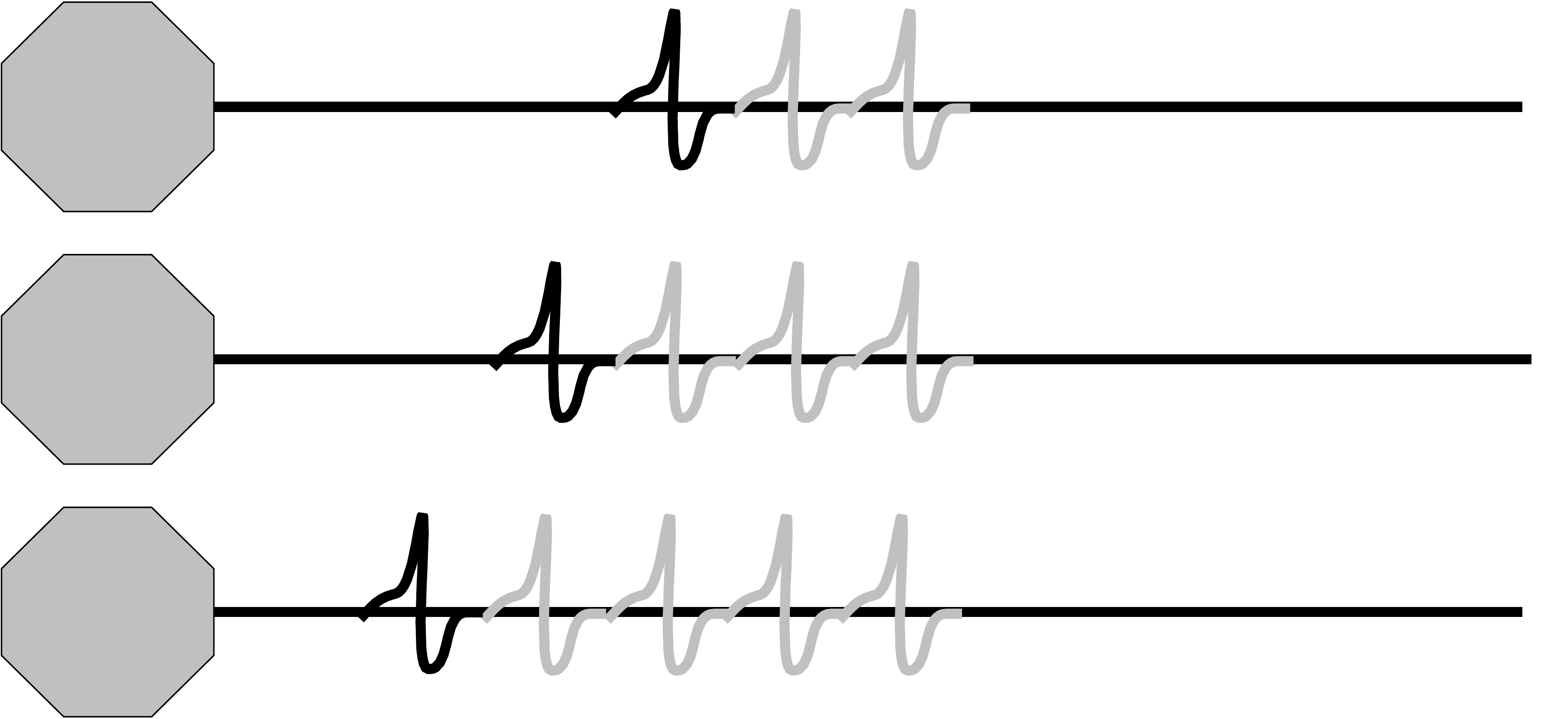}};
	
	\node [] (n1) at (4.5,0.85)   		 {$t2 = 8^0$};
	\node [] (n1) at (4.5,-0.1) 	 {$t1 =8^1$};
	\node [] (n1) at (4.5,-1.05) 	 {$t0 =8^2$};

	\draw [thick, ->] (arrow-begin) edge (arrow-end)
	node [pos=0.5,anchor=north,yshift=-1.75cm,xshift=0.6cm] {time};
	
	\draw [
	thick,
	decoration={
		brace,
		mirror,
	},
	decorate
	] (fix-begin.south) -- (fix-end.south)
	node [pos=0.5,anchor=north,yshift=-0.2cm] {
		$\begin{aligned}
		&(00011111\quad00001111\quad00000111)_{tr8}\\
		=&(8^2\times 5) \quad + (8^1\times 4) \quad+ (8^0\times 3) \\
		= &320 + 32 + 3 = (355)_{10}
		\end{aligned}$};
	
	\end{tikzpicture}
\vshrink{0.4}
	\caption{A value of $(101100011)_2$ = $(355)_{10}$ represented in base 8 temporal-rate encoding. Each neuron is assigned a weight relative to its spiking time. Then, each weight has a coefficient associated with the rate. The value is a weighted sum of each neuron.}
	\vshrink{0.6}	
	\label{fig:unary-positional-rate-temporal}
\end{figure}

This hybrid temporal-rate encoding where both spike timing and rate incorporated into the unary positional representation provides a sweet-spot among compactness, latency, resolution and error-tolerance, and avoids the drawback of temporal encoding in higher bases (i.e., it allows reliable comparison of values, so better accuracy for fine-grained classification or clustering). Figure~\ref{fig:unary-positional-rate-temporal} illustrates a temporal-rate encoding (in base 8) of a  value of $(101100011)_2$ as $(00011111\;00001111\;00000111)_{tr8}$ (where subscript $tr8$ indicates base 8 temporal-rate encoding). Each neuron is assigned a weight relative to its spiking time. Then, each weight has a coefficient associated with spiking rate of a neuron. The value is a weighted sum of each neuron. By temporal-rate encoding, a value can be represented with full accuracy in higher bases (as opposed to temporal encoding), while preserving compactness. 

\section{Conclusion and Future Work}
In this paper, we discuss the details of unary positional encoding that reduces the latency and length of the encoded streams (i.e., it is compact), and is error tolerant. We, then take a look at its prospect in spiking neural systems by incorporating temporal and rate encoding. 

As a future work, we would like to investigate on the complexity of building logic that
can operate on information represented in temporal-rate encoding (e.g., addition and multiplication) compared to traditional positional and unary encoding.


\bibliographystyle{ieeetr}
\bibliography{ref}

\end{document}